\documentclass[amssymb,aps,twocolumn,floats]{revtex4}
\usepackage{psfig}
\usepackage{color,graphicx,pstricks}
\draft
\begin{document}

\title{The Effect of Disorder in an Orbitally Ordered Jahn-Teller Insulator}

\author{Sanjeev Kumar$^{1}$, Arno P. Kampf$^{1}$, and Pinaki Majumdar$^{2}$ }

\affiliation{$^{1}$~Institute of Physics, Theoretical Physics III, Center for Electronic
Correlations and Magnetism,\\
University of Augsburg, D-86135 Augsburg, Germany\\
$^2$~Harish-Chandra  Research Institute,
Chhatnag Road, Jhusi, Allahabad 211 019, India
}

\begin{abstract}
We study a two dimensional, two-band double-exchange model
for $e_g$ electrons coupled to Jahn-Teller distortions
in the presence of quenched disorder
using a recently developed Monte-Carlo technique.
In the absence of disorder the half-filled system at low temperatures is an
orbitally ordered ferromagnetic insulator with a staggered pattern
of Jahn-Teller distortions. We examine the finite temperature transition to 
the  orbitally disordered phase and uncover a  qualitative difference 
between the intermediate and strongly coupled systems,
including a thermally driven insulator to metal 
crossover in the former case.
Long range orbital order is suppressed in the presence of disorder 
and the system displays a tendency towards metastable states consisting of
orbitally disordered stripe-like structures enclosing orbitally 
ordered domains. 

\vskip 0.2cm


\vskip 0.2cm

\today

\end{abstract}

\maketitle

\section{Introduction}

The undoped perovskite manganites, for example LaMnO$_3$,
are orbitally ordered antiferromagnetic insulators 
at low temperature. The magnetic order is of A-type with
ferromagnetic planes coupled antiferromagnetically in
the transverse direction. The orbital order is of C-type,
anti-correlated with the magnetic order \cite{biotteau01},
and is accompanied by long range ordering of the 
Jahn-Teller (JT) distortions of the MnO$_6$ octahedra. 
Upon hole doping, e.g. in La$_{1-x}$Ca$_{x}$MnO$_3$, the system evolves 
via a ferromagnetic insulating phase towards an 
orbitally disordered (OD) ferromagnetic metal, which
exhibits colossal magnetoresistance (CMR) near the Curie 
temperature \cite {vanaken03}.
Quenched disorder has been widely acknowledged as a crucial ingredient
for the understanding of hole-doped manganites \cite{attfield, tomioka}.
It is the source of a variety of phenomena,
both in real materials as well as in model calculations, like the
coexistence of metallic and insulating phases \cite{uehara,fath,moreo,nano,sk-apk-pm}, 
thermally driven metal to insulator 
transitions \cite{akahoshi,brey,dhde,sen06}, and
the suppression of charge order in half-doped systems \cite{motome,alvarez}.

Although the `fame' of the manganites rests on the 
CMR observed near optimal hole doping ($x \sim 0.3$) the low doping materials 
also display a rich variety of phenomena \cite{dagotto_book,tapan_book}. 
Hole doping leads to progressive loss of orbital order (OO),  
a reduction of the transport gap, increasing isotropy in the effective 
magnetic exchange,
and an insulator (I) to metal (M) transition \cite {vanaken03,palstra,hennion05}. 
Beyond transport and thermodynamic indicators,
NMR and neutron scattering experiments
have revealed an inhomogeneous - possibly
phase separated - state in the low doping regime \cite{papa03, hennion98}.
The interplay of doping, disorder, and thermal fluctuations 
thus presents an intriguing problem even in the insulators.
In typical hole doped manganite compounds such as RE$_{1-x}$AE$_x$MnO$_3$,
where RE (AE) denotes a trivalent (divalent) rare-earth (alkaline-earth) ion, 
the disorder arises from the difference in the
ionic radii of the RE and AE ions and hence, is tied to a
variation in the hole
concentration, vanishing for the stoichiometric compositions at $x=0$ and $x=1$. 
Conceptually, however, it is useful to
disentangle the effects of increasing hole density
and increasing disorder on the
OO-OD and IM transitions.
In that spirit, this paper focuses on 
the effect of disorder and varying electron-lattice coupling
at $x=0$ in a two-orbital model for the Mn $e_g$-electrons 
coupled to JT lattice distortions and to the 
$t_{2g}$-derived core spins.
The combined effect of disorder and hole doping has been 
briefly reported elsewhere \cite{sk-apk-pm}. 
A model system, which includes disorder remaining at
$x=0$ is experimentally relevant for systems 
with isoelectronic substitutions such as RE$_{1-y}$RE$'_y$MnO$_3$ \cite{bhattacharya04}.

The manganites are complex materials involving the 
the interplay of  spin, orbital, lattice, and charge degrees of freedom.
In the cubic perovskite structure, the 5-fold degeneracy of the Mn-$d$ levels is 
lifted by crystal fields leading to
two manifolds of three $t_{2g}$ and two $e_g$ orbitals. 
The energetically low-lying $t_{2g}$ levels contain three electrons and give
rise to a well localized $S=3/2$ spin. The $e_g$ electrons 
are itinerant and interact with
the distortions of the MnO$_6$ octahedra via the JT coupling and
with the core spins via Hund's rule coupling \cite{dagotto_book}.
The full complexity of these interactions
is hard to handle in a realistic two or three dimensional situation.
Let us quickly survey the attempts to understand the
$x=0$ state before describing our approach.

Different points of view have been reported regarding the relative importance of
the inter- and intra- orbital 
Hubbard interactions and the JT interaction in the manganites 
\cite{okamoto, millis-mueller}. 
Although the electron-electron (e-e) interactions are large in these systems,
it has been suggested that the parent insulator can not be considered as
a canonical Mott insulator \cite{held}.
Surprisingly, the orbitally ordered insulating groundstate at $x=0$ can be understood from a variety of starting points. For strong Hubbard interactions a spin-orbital t-J model
has been derived and studied within mean-field approximation \cite{kugel, ishihara}
leading to an orbitally ordered groundstate. 
The orbitally ordered groundstate also emerges from
Hartree-Fock band-structure calculations
considering the Mn-$3d$ and O-$2p$ orbitals \cite{HF1, HF2}. 
Monte-Carlo studies of models including JT coupling and 
ignoring the local Coulomb repulsion also
explain the OO-I state and tend to describe the system as a JT
insulator \cite{dagotto99}. In fact  
e-e and JT interactions are known to have
qualitatively similar consequences and instead of competing 
they re-enforce each other \cite{JT-ee}. 
More recently it was shown that a Fermi surface nesting instability 
at weak coupling can
also be the source for
an orbitally ordered groundstate in the presence of Hubbard interactions
\cite{efremov}. It is also known that a more realistic modelling of the
lattice taking into acount the individual oxygen displacements 
of MnO$_6$ octahedra and their cooperative effects, can by itself lead to a 
staggered JT distorted, and hence orbitally ordered, groundstate \cite{ahn}.
Retaining all the complications of the real materials is a difficult task,
particularly when we wish to advance beyond mean field theory.
The principal  purpose of the present study instead, is to
clarify the influence of 
thermal fluctuations and disorder on the orbitally ordered insulating groundstate.
Here we ignore the e-e interactions and the cooperative character of the
lattice distortions. The similar effects of e-e and JT interactions
in determining the nature of the groundstate suggests
that the effect of disorder
may also be similar for the two cases.

The remainder of the paper is organized as follows. 
In Section II we specify the model and 
briefly describe our Monte-Carlo simulation technique.
The results are presented in Sec. III, starting with the non-interacting system
and subsequently introducing the JT coupling and quenched disorder.
The non disordered zero temperature 
limit is independently analyzed by using variational calculations,
while the clean strong coupling limit studied via 
an effective classical model.

\section{Model and Method }

We consider a two-band model for itinerant $e_g$ electrons on a square lattice.
The electrons are coupled to JT lattice distortions,
$t_{2g}$ derived $S = 3/2$ core spins and quenched disorder
as described by the Hamiltonian:

\begin{eqnarray}
H &=& \sum_{\langle ij \rangle \sigma}^{\alpha \beta}
t_{\alpha \beta}^{ij} 
\left ( c^{\dagger}_{i \alpha \sigma} c^{~}_{j \beta \sigma} + h.c. \right )
+ \sum_i (\epsilon_i -\mu)n_i \cr
&&
~ - J_H\sum_i {\bf S}_i.{\mbox {\boldmath $\sigma$}}_i 
 - \lambda \sum_i {\bf Q}_i.{\mbox {\boldmath $\tau$}}_i 
+ {K \over 2} \sum_i {\bf Q}_i^2. ~ ~ ~ ~ ~ ~ ~ ~  
\end{eqnarray}

\noindent
Here, $c$ and $c^{\dagger}$ are annihilation and creation operators for
$e_g$ electrons and
$\alpha$, $\beta $ are summed over the two Mn-$e_g$ orbitals
$d_{x^2-y^2}$ and $d_{3z^2-r^2}$, which are labelled $(a)$ and $(b)$ in what follows.
$t_{\alpha \beta}^{ij}$ denote the hopping amplitudes between
$e_g$ orbitals on nearest-neighbor sites and have the cubic perovskite specific form:
$t_{a a}^x= t_{a a}^y \equiv t$, 
$t_{b b}^x= t_{b b}^y \equiv t/3 $,
$t_{a b}^x= t_{b a}^x \equiv -t/\sqrt{3} $,
$t_{a b}^y= t_{b a}^y \equiv t/\sqrt{3} $, where
$x$ and $y$ mark the spatial directions \cite{dagotto99}.
Disorder is modelled by random on-site potentials $\epsilon_i$
with equally probable values $\pm \Delta$, which
couples to the local electronic density $n_i$.
The $e_g$-electron spin is locally coupled to the
$t_{2g}$ spin ${\bf S}_i$ via the Hund's rule coupling 
$J_H$.
The electronic spin is given by ${\sigma}^{\mu}_i= 
\sum_{\sigma \sigma'}^{\alpha} c^{\dagger}_{i\alpha \sigma} 
\Gamma^{\mu}_{\sigma \sigma'}
c_{i \alpha \sigma'}$, 
where $\Gamma^{\mu}$ are the Pauli matrices.
$\lambda$ denotes the strength of the JT coupling between the distortion 
${\bf Q}_i = (Q_{ix}, Q_{iz})$ and  
the orbital pseudospin
${\tau}^{\mu}_i = \sum^{\alpha \beta}_{\sigma}
c^{\dagger}_{i\alpha \sigma} 
\Gamma^{\mu}_{\alpha \beta} c_{i\beta \sigma}$.
$K$ is a measure of the lattice stiffness, and $\mu$ is the chemical potential.
We set $t=1=K$ as our reference energy scale. 
The JT distortions and the $t_{2g}$ derived core spins 
are treated as
classical variables, and we set $\vert {\bf S} \vert =1$.
The present study is restricted to half-filling and we explore the   
variation in the parameters $\lambda$ and $\Delta$
in addition to the temperature $T$. 

For further simplification we adopt the limit $J_H/t \rightarrow \infty$, 
which is justified and frequently used in the context of manganites \cite {moreo, brey}. 
In this limit
the electronic spin at site $i$ is tied to the orientation 
of the core spin ${\bf S}_i$. Transforming the fermionic operators to this
local spin reference frame leads to the following 'spinless' 
model for the $e_g$ electrons:
\begin{eqnarray}
H &=& \sum_{\langle ij \rangle }^{\alpha \beta}
\left ( {\tilde t}_{\alpha \beta}^{~ij}
 c^{\dagger}_{i \alpha } c^{~}_{j \beta } + h.c. \right )
+ \sum_i (\epsilon_i -\mu)n_i \cr
&& ~ ~ ~ ~
 - \lambda \sum_i {\bf Q}_i.{\mbox {\boldmath $\tau$}}_i 
+ {K \over 2} \sum_i {\bf Q}_i^2.
\end{eqnarray}
The new hopping amplitudes have an
additional dependence on the core spin configurations and are given by:

\begin{eqnarray}
\frac{\tilde t_{\alpha \beta}}{t_{\alpha \beta}} = \cos \frac{\theta_i}{2}
\cos \frac{\theta_j}{2}
+\sin \frac{\theta_i}{2} \sin \frac{\theta_j}{2} ~ e^{-{\rm i}(\phi_i-\phi_j)}.
\end{eqnarray}

\noindent
Here, $\theta_i$ and $\phi_i$ denote polar and azimuthal angles for the spin ${\bf S}_i$.
From now on the operator $c_{i \alpha}$ ($c^{\dagger}_{i \alpha}$) is associated with 
annihilating (creating) an electron at site $i$, in the orbital $\alpha$ with 
spin parallel to~${\bf S}_i$.

The model given by Eq. (2) is bilinear in the electronic operators and 
does not encounter the problem of exponentially growing Hilbert space, since
all many-particle states can be constructed from Slater determinants of the 
single-particle states. The difficulty, however, arises 
from the large phase space in the classical variables ${\bf Q}$ and ${\bf S}$.
At zero temperature the problem reduces to finding the spin and lattice
configurations $\{ {\bf S}_i, {\bf Q}_i \}$ that minimize the total energy.
The energies for a limited number of periodic structures in 
${\bf Q}$ and ${\bf S}$ can be compared analytically.
This method is not assured, though, to lead to the true groundstate and often requires 
additional physics insight for identifying periodic structures, which 
are prime candidates
for the groundstate.
The finite temperature properties are not accessible in this manner, 
since they necessarily require the
electronic energies and wavefunctions in non-periodic 
structures for the ${\bf Q}$ and ${\bf S}$ variables. Further complications arise
in the presence of disorder and even the groundstate may not belong to
the subspace of periodic configurations of the spin and lattice variables.
Two controlled methods are available at finite $T$:
$(a)$~ dynamical mean-field theory (DMFT) correctly captures the 
strong coupling physics \cite{held00, millis-m-s} but is unable to handle 
spatially correlated 
inhomogeneous states, which may emerge in the presence of quenched disorder,
while, $(b)$~ exact diagonalization based Monte Carlo (ED-MC) calculations
\cite {dagotto_book}
can handle both thermal
fluctuations and disorder, but they are severely size limited. 

Here, we will use a scheme which is closely related to ED-MC
and employ the travelling cluster approximation (TCA)
\cite{tca}
in order to overcome the small size limitations. This approximation
is based on the observation, that the effect of a local update on
spin and lattice variables does
not `propagate' long distances via the electrons, and the
energy change involved in such a move can be computed by
constructing a cluster Hamiltonian around the reference site.
This drastically reduces
the computational cost and allows access to lattice sizes of $\sim 1000$ sites.
For the electronic properties in the thermally equilibrated 
system we use ED for the full system.
The details of this computational scheme have been  discussed previously \cite{tca}.
While TCA forms the backbone of the present study and is used
to map out the phase diagram of our model Hamiltonian 
in the parameter space of $\lambda$, $\Delta$, and $T$ with
detailed real-space information, we also analyze
the limiting cases using variational calculations,
or an effective classical model for the JT lattice distortions. Apart from 
providing a comparison to the TCA results, this also enables us to
establish a more transparent physical picture of the numerical results. 

\section{Results}

\subsection{Orbital order in the groundstate}

We begin by describing the simplest limit of the Hamiltonian in Eq. (2),
which is the non-interacting electron system in the absence of disorder.
Since the magnetism is purely double-exchange driven, 
leading to a ferromagnetic (FM) groundstate,  
a fully polarized core spin state is used for the calculation
of the electronic dispersion relation.
For $\lambda=0$ the spectrum is straightforwardly obtained by Fourier transformation
to momentum space:
\begin{eqnarray}
c_{i \alpha}^{\dagger} = \frac{1}{N} \sum_{\bf k} ~e^{ {\rm i} {\bf k} 
\cdot {\bf r_i}} d_{{\bf k} \alpha}^{\dagger}. 
~ ~ ~ ~ ~ ~,
\end{eqnarray}
This gives 
\begin{eqnarray}
H & = & \sum_{{\bf k}, \alpha \beta} 
\epsilon_{\alpha \beta}({\bf k}) ~ d_{{\bf k} \alpha}^{\dagger}  
d_{{\bf k} \beta}^{}
\end{eqnarray}
with $\epsilon_{\alpha \beta}({\bf k}) = -2 t_{\alpha \beta}^x cos(k_x)
- 2 t_{\alpha \beta}^y cos(k_y) $.
Diagonalizing this Hamiltonian we obtain

\begin{eqnarray}
 E_{ {\bf k}}^{\pm} = \frac{\epsilon^+({\bf k})} {2}
 \pm \sqrt { \left( \frac{\epsilon^-({\bf k})} {2} \right)^2 
 + \epsilon^2_{ab}({\bf k})} ,
\end{eqnarray}
where $\epsilon^{\pm}({\bf k}) = \epsilon_{aa}({\bf k}) \pm \epsilon_{bb}({\bf k}) $.

\begin{figure}
\centerline{
\includegraphics[width=8cm ,height = 8.7cm, clip=true]{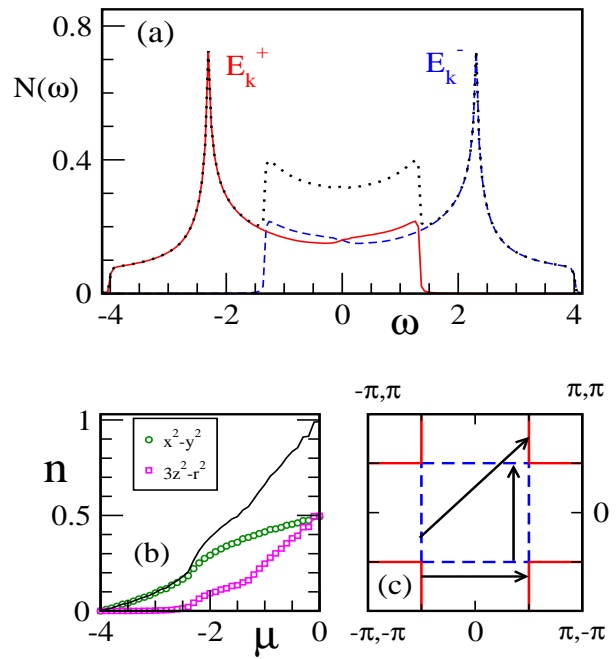}
}
\vspace{.1cm}
\caption{(Color online)
(a)~Density of states for the non-interacting Hamiltonian.
Solid (dashed) line is for the $E_{\bf k}^+$ ($E_{\bf k}^-$) band and the total
density of states is shown as the dotted line.
(b)~Filling of the $d_{x^2-y^2}$ and $d_{3z^2-r^2}$ orbitals
as a function of the chemical potential. The solid line is the total filling $n$. 
(c)~Fermi surfaces for the non-interacting model at $n=1$.
The solid (dashed) segment of the Fermi surface arises from the band $E_{\bf k}^+$ ($E_{\bf k}^-$).
The three nesting wave-vectors are indicated by the arrows.
}
\end{figure}

Fig. 1(a) shows the
band-resolved and the total density of states (DOS) corresponding to the dispersion 
given in Eq. (6). The electronic DOS is given by: 
$$
N(\omega) = \frac {1}{N} \sum_{n} \delta(\omega-E_{n})
$$
where $E_{n}$ denotes the eigenvalues of the Hamiltonian.
For the band-resolved DOS we use $E_{\bf k}^+$ or $E_{\bf k}^-$ in place of $E_{n}$.
Fig. 1(b) plots the filling of the $d_{x^2-y^2}$ and $d_{3z^2-r^2}$ orbitals as a function
of the chemical potential. These are simply the groundstate 
expectation values of the number
operators $d_{{\bf k} a}^{\dagger} d_{{\bf k} a}^{} $ 
and $d_{{\bf k} b}^{\dagger} d_{{\bf k} b}^{} $.

While the occupancy of the two orbitals is equal at $n=1$,
$d_{x^2-y^2}$ is preferred for $0 < n < 1$.
Fig. 1(c) shows the Fermi surface at electron filling $n=1$, which corresponds
to $\mu= 0$. From panel (a) it is clear that both bands are partially filled at
$\mu = 0$ and the Fermi surface consists of contributions from both bands.
While each of the bands $E_{\bf k}^{\pm}$
provides a segment of the Fermi surface, nested 
with wavevectors $(\pi,0)$ and $(0,\pi)$, there is 
also interband nesting with wavevector $(\pi,\pi)$.
The existence of the nesting wavevectors and a therefore
a divergent non-interacting susceptibility
induces an ordering instability in the presence of
interactions. For example, 
the inclusion of an infinitesimal electron-lattice coupling, 
$\lambda$, is expected to lead to a lowering
in the electronic energy by stabilizing the
lattice patterns, which are compatible with a selected
nesting wavevector. Similar arguments have been invoked earlier for the
same band dispersion to show
that the presence of weak e-e interactions leads to 
orbital ordering at half filling \cite{efremov}. 

In order to confirm this in the context of electron-lattice interactions, 
we perform a variational calculation
to search for the lowest energy state within a restricted set of possible groundstates.
As discussed earlier, the $T=0$ problem amounts to finding the spin and lattice
configurations $\{ {\bf S}_i, {\bf Q}_i \}$, which minimize the total energy.
Since the groundstate is known to be ferromagnetic, we are 
left with the problem of determining the 
minimum-energy lattice
configuration $\{ {\bf Q} \}$ only, for which we 
set up a restricted  variational calculation.
We consider variational states of the type
$ {\bf Q}_i = {\bf Q}~e^{{\rm i} {\bf q} \cdot {\bf r}_i} $ 
and compare energies for ${\bf q} = (0,0),~(0, \pi),~(\pi,0)$ 
and $(\pi,\pi)$. The variational parameters are the magnitude
of the distortion $Q$
and the orientation of ${\bf Q}$ in the
$Q_x-Q_z$ plane parameterized by an angle $\zeta$, with $\tan(\zeta) = Q_x/Q_z$.
For this restricted set of lattice configurations 
the Hamiltonian matrix is reduced to a
$ 2\times 2$ or $4\times 4$ form, which is
diagonalized to evaluate
the total energy
and to construct an approximate $T=0$ phase diagram.
The matrix for the uniform distortions (${\bf q} = {\bf 0} $) is 
$$ 
\left( \begin{array}{cc}
\epsilon_{11}({\bf k})-\lambda Q_z  &  \epsilon_{12}({\bf k})- \lambda Q_x \\
\epsilon_{12}({\bf k})-\lambda Q_x & \epsilon_{22}({\bf k}) + \lambda Q_z
\end{array} \right) ~ ~ ~ ~ ,
$$
and the matrix for ${\bf q} = (0,\pi)$,$(\pi,0)$ or $(\pi,\pi)$ is 
$$ 
\left( \begin{array}{cccc}
\epsilon_{11}({\bf k}) &  \epsilon_{12}({\bf k}) & -\lambda Q_z & -\lambda Q_x \\
\epsilon_{12}({\bf k}) & \epsilon_{22}({\bf k})  & -\lambda Q_x & \lambda Q_z  \\
-\lambda Q_z & -\lambda Q_x & \epsilon_{11}(k+{\bf q})  & \epsilon_{12}({\bf k}+{\bf q}) \\
-\lambda Q_x & \lambda Q_z & \epsilon_{12}({\bf k}+{\bf q}) & 
\epsilon_{22}({\bf k}+{\bf q})  \\
\end{array} \right) .
$$

\begin{figure}
\centerline{
\includegraphics[width=8cm ,height = 8.8cm, clip=true]{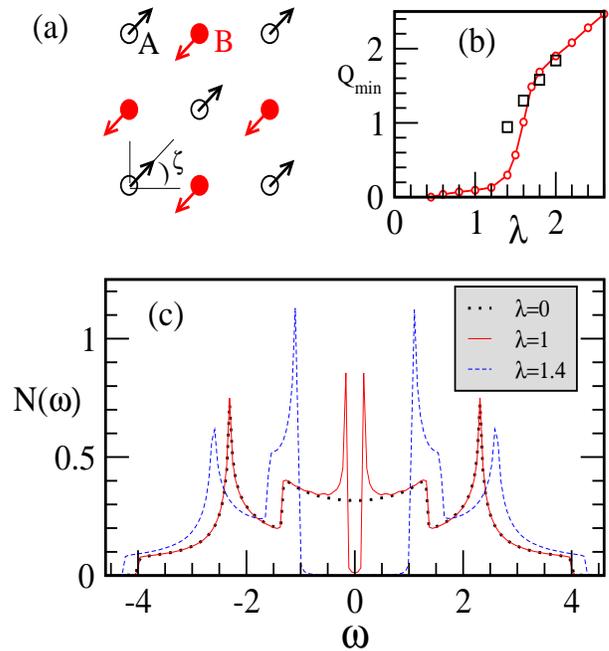}
}
\caption{(Color online)
(a)~Schematic picture of a staggered distortion pattern for the JT distortion vectors ${\bf Q}$.
    The arrows indicate the directions of ${\bf Q}$, with an arbitrary angle $\zeta$.
    The groundstate corresponds to $\zeta_A = \pi/2$, and $\zeta_B = 3\pi/2$.
 (b)~The variation of the magnitude of the distortion shown in (a) with $\lambda$.
     The circles are obtained from the variational calculation
     and the squares are from the Monte-Carlo simulations.
 (c)~Density of states for $\lambda = 0,~1$ and $1.4$  in the variational
     groundstate. Lattice sizes of up to $2000 \times 2000$ sites are used in the
     variational study.
}

\end{figure}

The results of the variational calculation are summarized in Fig. 2. 
A staggered pattern of JT distortions with an arbitrary
orientation angle $\zeta$ is schematically shown in Fig. 2(a). 
The minimum energy is obtained for $\zeta_A = \pi/2$ and $\zeta_B = 3\pi/2$.
The origin of purely $Q_x$-type distortion patterns is tied to the 
specific structure of the hopping parameters $t_{\alpha \beta}$
and especially to the sign difference between $t_{ab}^x$ and $t_{ab}^y$.
For $t_{\alpha \beta}^x \equiv t_{\alpha \beta}^y$ the minimum energy
configuration corresponds to a staggered distortion pattern
with $\zeta_A = \tan^{-1}[2 t_{ab}/(t_{aa} - t_{ab})]$ 
and $\zeta_B = \pi+\zeta_A$. For 
$t_{ab}^x = t_{ab}^y = t/\sqrt 3$ this leads to 
$\zeta_A = \pi/3$ and $\zeta_B = 4\pi/3$, and for $t_{ab}^x = t_{ab}^y = -t/\sqrt 3$ 
the corresponding angles are $2\pi/3$ and $5\pi/3$.
The relative sign difference between $t_{ab}^x$ and $t_{ab}^y$ in cubic 
perovskites,
thus leads to a conflict regarding the orientation angles for the
directions of ${\bf Q}$ on the two sublattices and $\zeta_A = \pi/2$, $\zeta_B = 3\pi/2$ 
result as a compromise.
The $Q_x$-type distortion patterns can also be motivated
from the structure of the effective classical model at strong JT coupling,
which is again related to the perovskite specific hopping parameters.
Since the lattice distortions are coupled to the orbital pseudospin,
$Q_x$-type ordering in the JT distortions is accompanied by
$\tau_x$-type staggered order in the orbital sector. 

Fig. 2(b) shows the the magnitude $Q_{min}$
of the distortion corresponding to the minimum energy state
as a function of the electron-lattice coupling.
For weak $\lambda$ there is a `BCS' like instability leading to 
an exponentially small $Q_{min}$ and a correspondingly
small gap in the DOS \cite{blawid}.
The window 
$0.0 < \lambda < 1 $ roughly corresponds to this weak coupling regime
and displays a slow   
rise in  $Q_{min}$ with $\lambda$.
For $1 < \lambda < 2 $, the intermediate coupling regime,
$Q_{min}$ increases rapidly
crossing over to the strong coupling asymptote, $Q_{min} \propto \lambda$,
 for 
$\lambda > 2$. The squares mark the data points obtained from TCA,
which are discussed later.
Fig. 2(c) shows the DOS for the lattice structure of panel (a) 
in the weak and intermediate coupling regime.

At strong coupling it is reasonable to start from the atomic limit and to assume that 
the electrons are site-localized by strong lattice distortions \cite{millis96}.
In that case the total energy per site, which includes the 
electronic and the elastic energy,
is given by
$$
E_{tot} = -\lambda|{\bf Q}| + {K \over 2} |{\bf Q}|^2 .
$$

Minimizing the total energy to obtain the optimum distortion $Q_{min}$,
one finds $Q_{min} = \lambda/K$. This is precisely the observed
behavior for $\lambda > 2$. Such a self-trapped object is usually
referred to as a polaron and the associated energy is termed
single-polaron energy $E_p = \lambda^2/{2K}$. In this paper, 
a 'polaron' always refers to a 'static polaron' since the
lattice is treated in the adiabatic limit.
The qualitative differences between intermediate and strong coupling
will become apparent in the finite temperature studies using TCA.

\subsection{Monte Carlo results}

\subsubsection{TCA in the clean limit}

The finite $T$ problem is solved using the TCA method described earlier. In this
case an unrestricted search is performed with respect to the spin and 
lattice configurations
and the temperature is reduced step by step to obtain the groundstate.
Fig. 3(a) shows the temperature dependence of the ${\bf q} = 
(\pi, \pi) \equiv {\bf q}_0 $ component of the lattice structure factor, 
$$ D_{Q} ({\bf q})~ = N^{-2} \sum_{ij} \langle
{\bf Q}_i \cdot {\bf Q }_j \rangle_{th} ~ ~ 
e^{-{\rm i}{\bf q}\cdot ({\bf r}_i - {\bf r}_j)}. $$
\noindent 
Here and below $\langle ... \rangle_{th}$ denotes the 
average over thermal equilibrium configurations.
All the TCA results are obtained on a $24 \times 24$
lattice and a $4 \times 4$ travelling cluster. 
$D_{Q} ({\bf q}_0)$ is a measure of the staggered
ordering tendency of the JT distortions.
Since the lattice distortions are coupled to the orbital 
pseudospin the same tendency is 
transferred to the analogously defined orbital structure factor 
$D_{\tau}({\bf q}_0)$. Therefore $D_{Q}({\bf q}_0)$ serves as an 
indicator for the staggered ordering of both, the lattice 
distortions and the orbital pseudospin. The point of inflection in the
temperature dependence of $D_{Q}({\bf q}_0)$ is taken as the 
orbital ordering transition temperature $T_{OO}$.
Fig. 3(b) shows the variation of $T_{OO}$
with $\lambda$. The symbols are TCA results and the solid line is a
guide to eye. 
The extrapolation of the solid line to $\lambda < 0.8$ is based on the
existence of a BCS-like instability at weak $\lambda$ due to Fermi-surface nesting
as already encountered in the variational calculations.
In the weak-coupling regime $T_{OO}$ is proportional to 
the energy gap $\Delta_G$ in the DOS. 
The dashed line indicates the result of a strong coupling expansion,
which leads to the proportionality $T_{OO} \propto t^2/E_p$.
The details of the strong coupling expansion are discussed in the appendix.
The non-monotonic behavior of $T_{OO}$ is thus naturally understood by merging the
weak and strong coupling limits.
The experimental results on REMnO$_3$ for the 
accessible range of decrease in mean ionic
radii show an increase in $T_{OO}$ \cite{kimura}.
A reduction in ionic radii leads to a decrease in the bandwidth,
which translates to an effective increase in $\lambda/t$ in our model.
This suggests that the relevant regime for the strength of JT coupling 
should be $\lambda < 1.8$. A more direct estimate for the value of
$\lambda$ is obtained by comparing the ratio $\Delta_G/T_{OO}$
between our results and the experiments. For LaMnO$_3$, $\Delta_G \sim 0.5eV$ 
\cite{palstra} and $T_{OO} \sim 700K $ \cite{kimura} leading to $ \Delta_G/T_{OO} \sim 8$.
This value is approximately reproduced for $\lambda \sim 1.2$.
In real materials additional interactions may lead to a renormalization 
of the effective JT coupling \cite{JT-ee},
therefore these numbers should be considered only as a rough estimate.

\begin{figure}
\centerline{
\includegraphics[width=8.4cm ,height = 4.8cm, clip=true]{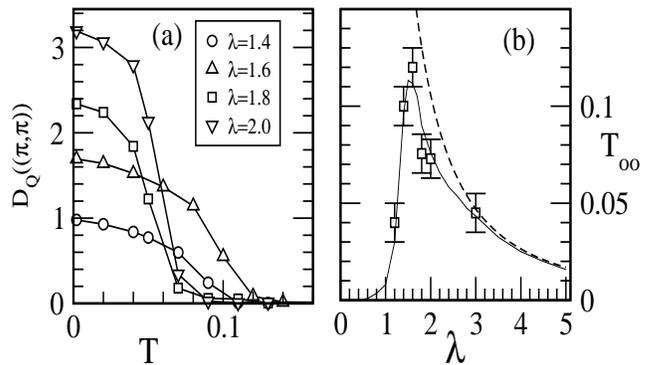}
}
\caption{(a)~ Temperature dependence of the lattice structure factor 
$D_Q((\pi,\pi))$. (b)~ Variation of the orbital-ordering transition temperatures
$T_{OO}$ with $\lambda$. The squares are the TCA results and the solid line is
a guide to eye. The strong coupling asymptotic form is also shown by the dashed line.
}
\vspace{.5cm}
\end{figure}

Since the orbital order is tied to the ordering of the JT
distortions, it is useful to track the temperature evolution of the
distribution function for the lattice distortions
$$
  P(Q) = \frac {1}{N} \left \langle \sum_{i} \delta(Q - |{\bf Q}_i|) \right \rangle_{th} ~,
$$
where $|{\bf Q}_i|$ denotes the magnitude of the local lattice distortions 
for an equilibrium configuration.
The temperature dependence of $P(Q)$ is shown in Figs. 4(a) and 4(b)
for moderate and strong JT coupling strengths, respectively.
The low $T$ distributions are peaked at $Q = Q_{min}$, suggesting an unimodular
structure at $T=0$ and supporting the variational results for the groundstate.
The symbols represent the TCA data and the solid lines follow from
simple arguments for the expected structure of the distribution function $P(Q)$:
The single-site energy for an electron self-trapped by
a distortion of magnitude $Q$ is given by $E = -\lambda ~ Q
+ K/2 ~ Q^2$. The probability that this site has a distortion with magnitude
$Q$ at a given temperature $T$ is then given by
$P(Q) \propto exp[~-(E-E_0)/T]$, where $E_0$ is the groundstate
energy correspnding to $Q = Q_{min}$. This leads to
$P(Q) \propto exp[~-(Q-Q_{min})^2/2T]$, assuming that the
functional form of the energy does not change for finite $T$.
Using the $Q_{min}$ data of the
TCA results these functions are plotted as solid lines in Figs. 4(a) and 4(b).
The TCA results show that, $P(Q=0)$ is finite for $T \geq T_{OO}$ at $\lambda = 1.4 $; 
thus some of the electrons can apparently escape from JT-distorted cages upon heating.
This suggests the possibility of an insulator to metal crossover 
and will 
be discussed in the next section.
The naive analysis described above does not apply for weak coupling,
as is apparent from the large deviations of the firm lines from the symbols
, infact, even the width at low temperature of the distribution function is not captured
by the corresponding Gaussian function.
For $\lambda=2$ however, the electrons are well 
trapped by the self-generated JT distortions even at 
$T \sim 3 T_{OO}$. The Gaussian functions 
with width $\propto \sqrt T$ are reasonable fits to
the TCA data, implying that $\lambda=2$ is close to the regime where single-site
analysis is valid.
This clarifies a crucial difference between the weak and the strong
coupling systems despite the fact that the orbital order in the groundstate is 
the same in both limits. In the strong coupling regime the distortions exist
even at high temperatures and the orbital ordering transition corresponds to the
alignment of the distortion vectors at $T \sim T_{OO}$.
This is analogous to the ordering in a spin model with magnetic moments of
fixed magnitude. The mechanism for orbital ordering at weak to moderate coupling
relies on a simultaneous generation and ordering of the lattice distortions.

\begin{figure}
\centerline{
\includegraphics[width=8.2cm ,height = 8.4cm, clip=true]{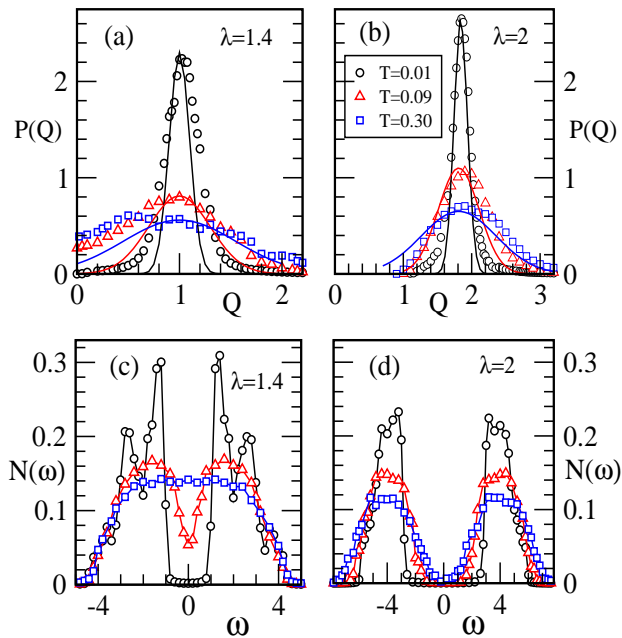}
}
\caption{ (Color online)~ Temperature evolution of the distribution function for 
the magnitude of the lattice
distortions $P(Q)$ for moderate (a), and strong (b) electron-lattice
coupling strengths. Symbols are the Monte-Carlo data and the solid lines 
correspond to the naive expectation based on the atomic limit.
The density of states for the parameters corresponding to 
(a) and (b) is shown in (c) and (d), respectively. 
}
\end{figure}

Although the stability of these results has been checked for system sizes 
ranging from $N = 8^2$ to $N = 32^2$,
a detailed finite size scaling has not been performed. The observed qualitative
difference between the moderate and strong coupling systems raise the possibility,
that the orbital order to disorder transitions in the two cases may
be qualitatively different.
Experimentally it is known that the transition in LaMnO$_3$ is 
first order in nature and becomes second order for PrMnO$_3$ and NdMnO$_3$,
which are smaller bandwidth materials \cite{bhattacharya04, zhou, chatterji}.
Although the origin for a first order nature of the transition in LaMnO$_3$
is suggested to be anharmonic coupling between JT distortions and volume strain \cite{maitra},
the bandwidth variation may also be playing a role.
Figs. 4(c) and 4(d) show the electronic DOS for the same parameter values as used in
(a) and (b), respectively. The low $T$ DOS is gapped, 
consistent with our variational results.
The gap fills up for $T \sim T_{OO}$ for weaker $\lambda$ but persists to much larger
$T$ for strong coupling. 
The origin of
the gap in the DOS is therefore very different for the two cases and finds supports in the
structure in $P(Q)$. Indeed, if the distortions are well formed even at
large $T$, as is the case for strong coupling, the DOS is gapped due to 'self-trapping'.
On the other hand a gap in the DOS for moderate coupling 
is tied to the existence of orbital order, which 
in turn requires a long range ordered pattern for the lattice distortions.
Note that the sizes of both $Q_{min}$ and 
the gap in the DOS obtained in the TCA calculation 
match very well with the variational calculations (see Fig. 2(b)).

\subsubsection{The effect of disorder}
Although the presence of disorder is usually tied to
the hole-doped materials, it can also be realized in compositions like
RE$_{1-y}$RE$'_y$MnO$_3$. Moreover it is useful to study the effect of
disorder without involving the complications of finite hole density.
Therefore, given the presence of orbital order in the half-filled clean system, 
we now include the effects of quenched disorder. 
Although the Monte-Carlo results provide us with the full finite-$T$ information,
we first present our results on the disorder effects at low $T$. In this section
we focus on the thermodynamic quantities, which are 
averaged over $\sim 10$ realizations of quenched disorder.
Fig. 5(a) shows the staggered component of the 
lattice structure factor at low $T$ as a function of 
the disorder strength $\Delta$. 
The critical value $\Delta_c$ for the disappearence of
orbital order in the groundstate is estimated from these data and
further confirmed by the $T$ dependence of $D_Q({\bf q}_0)$.
$D_Q({\bf q}_0) \sim O(1)$ is considered to 
indicate orbital order and $D_Q({\bf q}_0) \sim O(1/N^2)$
implies an orbitally disordered phase.
A non-monotonic dependence of $\Delta_c$ on the JT coupling strength is observed as 
shown in Fig. 5(b).
The similarity between $\Delta_c(\lambda)$ 
and $T_{OO}(\lambda)$ (see Fig. 3(b)) suggests that the
effects of quenched disorder and thermal fluctuations are similar 
in  weakening the long range orbital order.

\begin{figure}
\centerline{
\includegraphics[width=8.4cm ,height=5.2cm, clip=true]{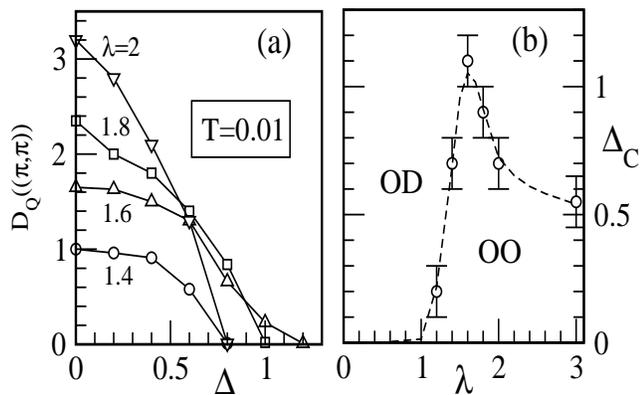}
}
\caption{
(a)~Lattice structure factor $D_Q((\pi,\pi))$ at $T=0.01$ as a function of 
the disorder strength $\Delta$. 
(b)~Critical disorder strength $\Delta_c$ required to spoil the orbital order in the
    groundstate as a function of $\lambda $.
}
\end{figure}

For further details on the disorder driven orbital order to disorder crossover,
we show the disorder evolution of the distribution function $P(Q)$ for the JT distortions 
for two representative values of $\lambda$ in Fig. 6(a) and 6(b). 
These results are very similar to the $T$ dependence of 
the distribution function (see Figs. 4(a)-(b)). Interestingly disorder
acts as a delocalizing agent for the weak coupling system, leading to a fraction
of sites with very weak JT distortions, hence delocalizing a fraction of electrons from the 
self generated JT traps. The effect of the disorder on $P(Q)$ is barely visible at large 
$\lambda$ and $P(Q)$ is not affected in crossing over from the orbitally ordered
to the orbitally disordered state.
Figs. 6(c) and 6(d) show the effect of disorder on the DOS for the same values of
parameters as used in panels (a) and (b). Our earlier suggestion that the gap in the DOS
is related to the orbital ordering for weak electron-lattice coupling is 
confirmed. For $\Delta = 0.8$ a pseudogap feature replaces the clean gap in the DOS.
The strong coupling DOS is only slightly affected by disorder and in particular the
gap survives even in the orbitally disordered phase. This is again similar to the 
temperature dependence of the DOS for the weak and strong coupling limits.

\begin{figure}
\centerline{
\includegraphics[width=8.2cm ,height = 8.6cm, clip=true]{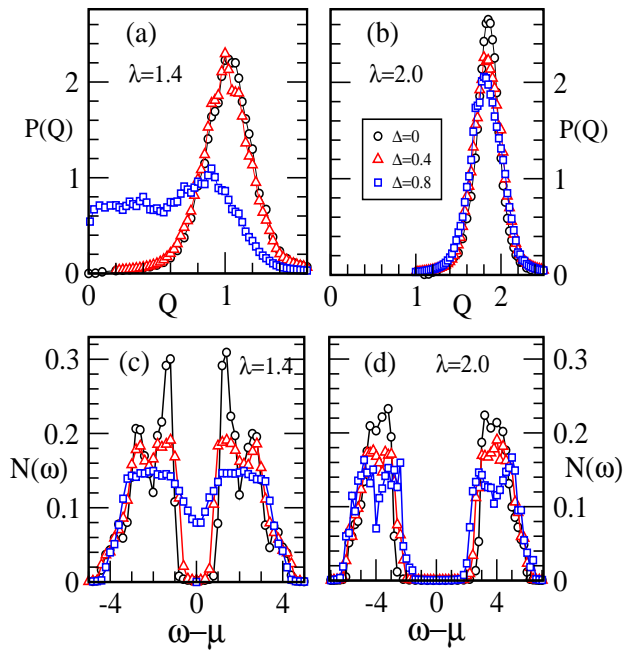}
}
\caption{ (Color online)~ Disorder dependence of the distribution function
$P(Q)$ for moderate (a), and strong (b) electron-lattice
coupling strengths. The density of states for the parameters corresponding to 
(a) and (b) is shown in (c) and (d), respectively. 
}
\end{figure}

The effect of disorder is also different for the finite $T$ properties
between weak and strong coupling.
Fig. 7 shows the temperature dependence of $D_Q({\bf q}_0)$
for various $\Delta$ at moderate (panel (a)) and strong (panel (b))
JT couplings. The $T=0$ value of $D_Q({\bf q}_0)$
reduces with increasing disorder in both cases. The detailed real-space
analysis of this reduction will be presented later, but it is
worth mentioning already here that orbitally disordered stripe-like structures emerge in the
disordered system at low temperatures, specially in the strong coupling regime. 
While the inclusion of disorder affects the 
saturation values for $D_Q({\bf q}_0)$ in a similar manner for both $\lambda = 1.4$
and $\lambda = 2.0$, 
$T_{OO}$ scales show an interesting contrast between the two coupling strengths.
$T_{OO}$ reduces
gradually for $\lambda = 1.4$ with increasing disorder strength and eventually both the
saturation value of $D_Q({\bf q}_0)$ and $T_{OO}$ approach zero as the
long range orbital order in the groundstate is lost. On the other hand, $T_{OO}$ does
not change with disorder for $\lambda = 2.0$ and drops abruptly to zero close to $\Delta_c$,
when the system crosses over to the orbitally disordered phase. 

For further confirmation on the difference between the weak and strong
coupling systems we compute the optical conductivity $\sigma(\omega)$ using 
the Kubo formula with the exact eigenstates 
\cite{transport}. The resistivity $\rho$ is approximated
by the inverse of $\sigma(\omega_{min})$, where $\omega_{min} = 20t/N
\sim 0.03t$ is the lowest reliable frequency 
scale for $\sigma(\omega)$ calculations on our $N=24^2$ system.
Fig. 7(c) shows the resistivity for $\lambda = 1.4$ as a function of $\lambda$
for different values of disorder strength. A sharp upturn in the resistivity
is observed upon cooling, which is clearly due to the onset of
orbital ordering. For all values of
$\Delta < \Delta_c$ a change of slope in the resistivity is observed, indicating a
thermally driven I-M crossover. For $\Delta \geq \Delta_c$, the system retains
$d\rho/dT > 0$ for all $T$, indicating a disorder-induced metallization of the insulator.
While the Anderson localization effects may not allow a metallic state in lower dimensions,
the I-M transition induced by quenched disorder via the destruction of orbital order
is expected in three dimensional systems too. The observation of a
disorder-induced metallic state is not new and has been reported before in
the context of charge ordering \cite{motome, alvarez}.
The temperature dependence of the resistivity for $\lambda=2$ is shown in
Fig. 7(d). This strong coupling system displays a more robust insulating behavior.
The effect of thermal fluctuations and/or disorder is barely visible on the
resistivity. Contrary to the weak-coupling system, 
the onset of orbital ordering is not reflected in $\rho(T)$.

We observe a monotonic decrease of $T_{OO}$ scales upon increasing disorder
in the experimentally relevant regime of moderate JT coupling. In real systems however,
it is difficult to isolate the individual effect of quenched disorder.
Isoelectronic rare-earth substitution series such as in 
La$_{1-y}$Nd$_y$MnO$_3$, involve a simultaneous 
change of effective bandwidth and disorder \cite {bhattacharya04}. 
The increase in $T_{OO}$ upon increasing $y$ in this series is understood mainly from the increase 
in the average ionic radius but changes in $y$ are necessarily accompanied by
variations in disorder.

\begin{figure}
\centerline{
\includegraphics[width=8.4cm ,height = 8.6cm, clip=true]{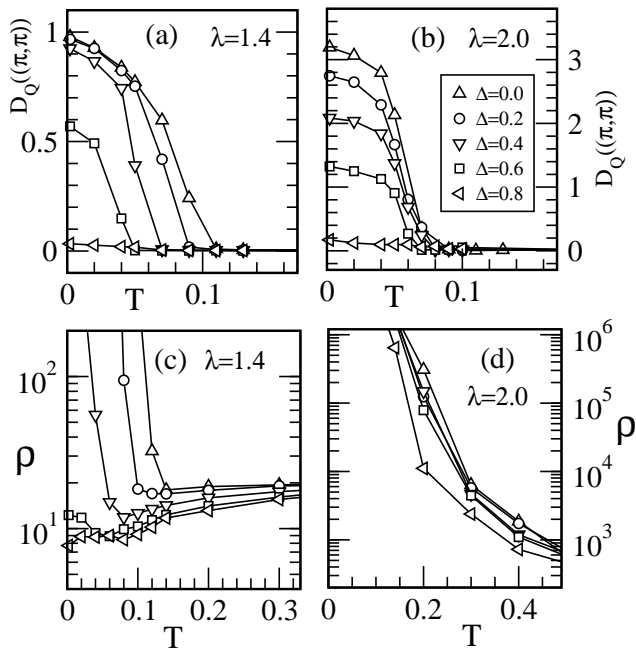}
}
\caption{~ Temperature dependence of the lattice structure factor 
$D_Q((\pi,\pi))$ for varying disorder strengths at 
(a) $\lambda = 1.4$ and (b) $\lambda = 2.0$. Resistivity $\rho$ in units of
$\hbar/\pi e^2$, as a function of
temperature for (c) $\lambda = 1.4$ and (d) $\lambda = 2.0$ for the same values of
$\Delta$ as in (a) and (b).
}
\end{figure}

One of the advantages of the Monte-Carlo method is the access to 
detailed real-space information. This is especially useful for determining
the mechanism leading to the loss of orbital order upon increasing
disorder strength. In Fig. 8 we show the local spatial correlations
of the JT distortions, as represented by the quantity $C_Q^i
=  \frac {1}{4} \sum_{\delta} {\bf Q}_i \cdot {\bf Q}_{i+\delta}$, where
$\delta$ is summed over the nearest neighbors of site $i$. Therefore 
$C_Q^i \sim -Q_{min}^2$ represents a region of staggered orbital ordering
while $C_Q^i \sim 0$ corresponds to an OD neighborhood.
Panels (a)-(c) show the groundstate pattern for $C_Q^i$ at moderate strength
of the JT coupling for three typical disorder realizations. In some cases 
orbitally disordered filamentary structures are observed.
These OD domain walls separate the orbitally ordered regions, and a $\pi$-phase shift
appears in crossing from one OO region to the other.
The patterns for $C_Q^i$ at strong coupling are shown in panels (d)-(f). Again
the data are presented for three typical realizations of quenched disorder.
The tendency to form filamentary structures of orbitally disordered regions is
even stronger for strong JT coupling. A careful analysis of these structures shows
that these are not 'unique' groundstates of the system and should be understood
as metastable states. Even for a fixed realization of disorder a different starting
configuration of the lattice variables in the Monte-Carlo simulations
leads to a different filamentary structure. Moreover,
in most cases these states are higher in energy as compared to the states
with no domain walls.
Upon doping the OO insulator, similar
stripe-like structures of OD regions appear as true groundstates and serve as
the prefered locations for the doped holes \cite{sk-apk-pm}. While the
earlier study involved a combined effect of quenched disorder and hole-doping,
here we propose that disorder itself 
leads to the existence of {\it metastable low energy configurations}
with 
filamentary OD regions enclosing OO domains. The disorder in our model couples to
the total charge density and in order to understand how the effect is transferred to
the orbital degrees of freedom, we analyze the effect of single impurity in the
orbitally ordered groundstate. Considering a repulsive impurity, the impurity site
leads to a slightly lower value of charge density. Since the band dispersion is such that
$n_a \neq n_b$ for $n \neq 1$, the sites with $n < 1$ has a finite value of $\tau^z$. These
finite $\tau^z$ values upset the purely $\tau^x$-type order. 
 
\begin{figure}
\centerline{
\includegraphics[width=8.4cm ,height = 6.0cm, clip=true]{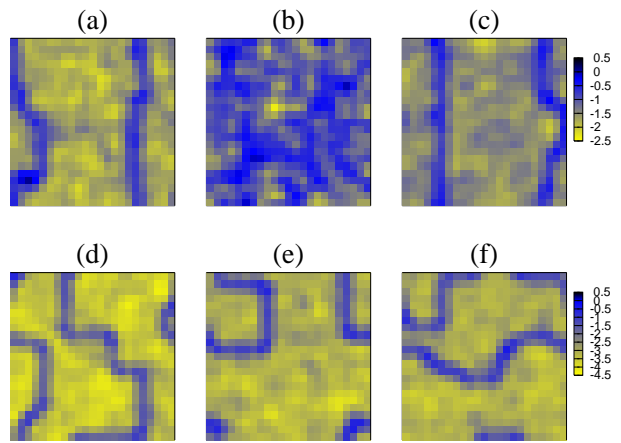}
}
\caption{ (Color online)~ Local spatial correlations $C_Q^i$ (see text) of the Jahn-Teller
distortions at $T=0.01$ for three typical realizations of
quenched disorder. Panels (a)-(c) correspond to $\lambda = 1.6$, $\Delta=0.5$, and 
panels (d)-(f) are for $\lambda = 2.0$, $\Delta = 0.6$. 
}
\end{figure}

\section{Conclusions}
In summary, we have studied a two-band double-exchange model at half filling for $e_g$ electrons
coupled to Jahn-Teller distortions in the presence of quenched disorder 
in two dimensions. The existence of long range orbital order
and the accompanying staggered order in the Jahn-Teller distortions in the
groundstate is verified by means of variational calculations, a strong coupling
analysis, and real-space Monte-Carlo simulations. The qualitative difference
between the moderate and strong Jahn-Teller coupling systems is reflected in the
nature of the orbital ordering transitions in the two regimes. A thermally
driven crossover from an insulator to a metal is observed for
weak to moderate coupling, in contrast to a robust insulating phase 
at strong coupling. Quenched disorder destablizes the orbitally ordered groundstate 
and leads to the appearance of metastabe states with orbitally disordered domain walls
separating the orbitally ordered regions. These metastable, inhomogeneous states
in the insulating half-filled system are important for the understanding of
structures and phase transitions in the hole doped systems 
\cite{sk-apk-pm}. 

\vspace{0.4cm}
\begin{center}
{\normalsize {\bf ACKNOWLEDGMENTS}}
\end{center}

SK and APK gratefully acknowledge support by the 
Deutsche Forschungsgemeinschaft through SFB 484.
PM was supported during part of this work by 
the EPSRC (UK) and Trinity College at Cambridge University, by the
Royal Society (UK) at Oxford University, and by the Institut 
Laue-Langevin at Grenoble.
Simulations were performed on the Beowulf Cluster at HRI.

\vspace{0.4cm}
\begin {center}

{\normalsize {\bf 
APPENDIX: EFFECTIVE CLASSICAL MODEL}}
\end{center}

In the limit of large JT coupling, the electrons are essentially site-localized
and the atomic limit ($t_{\alpha \beta} \equiv 0$) is a good starting point. 
The electronic eigenvalues
for a single-site problem are given by $\pm \lambda |{\bf Q}|$. Only the lower electronic
level is occupied for the half-filled case. Minimizing the
sum of electronic and lattice energies for the optimum value of the
magnitude of the JT distortion ${\bf Q}$ leads to $Q_{min} = \lambda/K$.
While the magnitude of the JT distortion is determined locally by
a balance between the elastic and the electronic energy, there
is no prefered direction for the orientation of ${\bf Q}$. This degeneracy is lifted
by including a finite hopping amplitude for electrons and leads to an effective classical
model for the orientations of the JT lattice vectors ${\bf Q}$. 
The derivation is straightforward and the main steps are outlined below:

\begin{figure}
\centerline{
\includegraphics[width=5.7cm ,height = 6.2cm, clip=true]{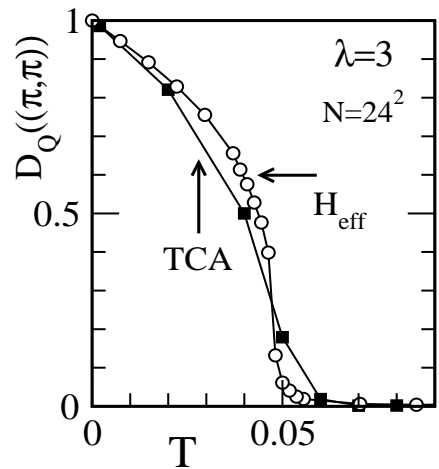}
}
\caption{~ Lattice structure factor $D_Q((\pi,\pi))$ as a function of
temperature for the effective strong-coupling model compared with the 
TCA results at $\lambda=3.0$. $D_Q((\pi,\pi))$ is normalized to its
saturation value for the TCA data.
}
\end{figure}

Consider a two site problem with $ H' = H_0 + V $
\begin{eqnarray}
H_0 &=& -\lambda \sum_{i=1,2} \{ Q_{ix} \tau_{i}^x + Q_{iz} \tau_{i}^z \}
        + \frac {K}{2}\vert{\bf Q}_i\vert^2 \\
V &=& \sum_{\alpha \beta} ( t_{\alpha \beta} ~ c_{1\alpha}^{\dagger}c_{2\beta} + h.c.  ) 
\end{eqnarray}

The eigenvalues of $H_0$ are $\pm \lambda \vert{\bf Q}_i\vert$ with $i=1,2$.
The eigenvectors $ \vert \psi_{\pm} \rangle$ are site-localized, and 
consist of  linear combinations  of
the two orbitals at a site. The eigenvector for the eigenvalue $-\lambda \vert{\bf Q}_i\vert $
($\lambda \vert{\bf Q}_i\vert $) corresponds to ${ \mbox {\boldmath $\tau$}}$ pointing parallel
(antiparallel) to ${\bf Q}$, which has the full rotational symmetry of the plane.
Explicitly, the eigenvectors are:

\begin{eqnarray}
 \vert \psi_- \rangle_i &=& \left \{ \cos \left({\frac{\zeta_i}{2}} \right ) ~ c_{ia}^{\dagger}
 + \sin \left( {\frac{\zeta_i}{2}} \right )c_{ib}^{\dagger} \right \}|0 \rangle , \\
 \vert \psi_+ \rangle_i &=& \left \{ -\sin \left ( {\frac{\zeta_i}{2}} \right )  ~ c_{ia}^{\dagger} 
 + \cos \left ( {\frac{\zeta_i}{2}} \right )c_{ib}^{\dagger} \right \} |0 \rangle ,
\end{eqnarray}
where $\zeta_i = \tan^{-1}(Q_{ix}/Q_{iz})$ denotes the orientation of the JT distortion vector ${\bf Q}_i$ 
in the $Q_{x}$-$Q_{z}$ plane.

\noindent
 The total energy of this 2-site system is
\begin{eqnarray}
 E = - \lambda\sum_{i=1,2} \vert{\bf Q}_i\vert + \frac {K}{2}\vert{\bf Q}\vert_i^2  ~ ,
\end{eqnarray}
which is minimized by $ \vert{\bf Q}_i\vert = \lambda/K $ independent of $i$. 
The two electron groundstate of $H_0$ thus corresponds to having one electron at each of the two sites
with a lattice distortion $\vert{\bf Q}\vert = \lambda/K$, total energy
$
 E_0 = - \frac {\lambda^2}{K}
$ and the groundstate wavefunction $
\vert \psi_0 \rangle = \vert \psi_- \rangle_1 \otimes \vert \psi_- \rangle_2$.

 Turning on the hopping part of the Hamiltonian $V$, the perturbative correction
 to the energy is given by
\begin{eqnarray}
\Delta E = \sum_{m} \frac { \left| \langle \psi_m | t_{\alpha \beta} 
(c_{1 \alpha}^{\dagger}c_{2 \beta}+h.c.)
| \psi_0 \rangle \right|^2}{E_0-E_m} ~ ~ ,
\end{eqnarray}
where, $|\psi_m \rangle$ denotes an excited state of $H_0$ in the two-electron subspace.
Explicit evaluation leads to

\begin{eqnarray}
\Delta E &=& J_x ~ \sin{\zeta_1}\sin{\zeta_2}
+ J_z ~ \cos{\zeta_1} \cos{\zeta_2} , \nonumber \\
& &
+ J_m ~ \left ( \sin{\zeta_1}\cos{\zeta_2}
+ \cos{\zeta_1}\sin{\zeta_2} \right ) ~ ~ ,
\end{eqnarray}
 with coupling constants
 $J_x = \frac{K}{\lambda^2} (t_{aa}t_{bb} + t_{ab}^2) ~$, \\
 $J_z = \frac{K}{\lambda^2} (\frac{(t_{aa}^2+t_{bb}^2)}{2} - t_{ab}^2) ~$, and 
 $J_m = \frac{K}{\lambda^2} [t_{ab} (t_{aa}-t_{bb})] ~$.

\noindent
This energy acts as an effective classical Hamiltonian for
the ordering of the distortion vectors ${\bf Q}$.

For a comparison with the results obtained from the TCA, we show
the temperature dependence of $D_Q({\bf q}_0)$ for the effective model and
TCA in Fig. 9. 
These results are obtained on the same lattice sizes as used for the TCA calculations, 
i.e. $24 \times 24$. The TCA curve is normalized by its $T=0$ value. The coefficients used in 
$H_{eff}$ have an additional factor of $1/2$ because
the system is still paramagnetic when the orbital ordering takes place and the
hopping amplitude $t$ has to be replaced by $t \sqrt {(1+{\bf S}_i.{\bf S}_j)/2}$,
which becomes $1/ \sqrt 2$ in the paramagnetic phase. The $T_{OO}$ scales obtained from
$H_{eff}$ match very well with the TCA scales
for $\lambda \geq 3$ and begin to deviate significantly only for $\lambda \leq 2$
(see Fig. 3(b)). The purely $Q_x$-type staggered ordering in the JT distortion vectors
is also observed within the effective classical Hamiltonian. This can be easily understood
by looking at the values of coupling constants $J_x$, $J_z$ and $J_m$. Again the difference
in sign between the values for $J_m^x$ and $J_m^y$, which originates from the sign difference
between $t_{ab}^x$ and $t_{ab}^y$, turns out to be crucial. 

{}

\end{document}